\documentclass[longauth]{aa} 
\newcommand{\HRI}{HRI$_{\rm EUV}$}
\newcommand{\HRIL}{HRI$_{\rm Lya}$}
\usepackage{booktabs}

\usepackage{graphicx}
\usepackage{subfigure}
\usepackage{txfonts}
\usepackage{hyperref}
\hypersetup{
    colorlinks,
    citecolor=blue,
    filecolor=blue,
    linkcolor=blue,
    urlcolor=blue}

\begin{document} 

   \title{Automatic detection of small-scale EUV brightenings observed by the Solar Orbiter/EUI}

   \author{N. Alipour \inst{\ref{i:ZN}},
        H. Safari\inst{\ref{i:ZN},\ref{i:OZN}}
          \fnmsep\thanks{Corresponding author: Hossein Safari \email{safari@znu.ac.ir}},
        C. Verbeeck \inst{\ref{i:Be}},
        D. Berghmans \inst{\ref{i:Be}}, 
        F. Auch\`ere \inst{\ref{i:Fr}}, 
        L. P. Chitta \inst{\ref{i:Ge}}, 
        P. Antolin \inst{\ref{i:UK}},
        K. Barczynski \inst{\ref{i:Sw1}, \ref{i:Sw}}, 
        \'E. Buchlin \inst{\ref{i:Fr}}, 
        R. Aznar Cuadrado \inst{\ref{i:Ge}},
        L. Dolla \inst{\ref{i:Be}}, 
        M.~K.~Georgoulis \inst{\ref{i:acad_at}},
        S. Gissot \inst{\ref{i:Be}},
        L. Harra \inst{\ref{i:Sw},\ref{i:Sw1}}, 
        A.~C.~Katsiyannis \inst{\ref{i:Be}}, 
        D.~M.~Long \inst{\ref{i:UK2}}, 
        S. Mandal \inst{\ref{i:Ge}}, 
        S. Parenti \inst{\ref{i:Fr}},
        O. Podladchikova \inst{\ref{i:Ge2}}, 
        E. Petrova \inst{\ref{i:Ru}},
        \'E. Soubri\'e \inst{\ref{i:Fr}}, 
        U. Sch\"uhle \inst{\ref{i:Ge}}, 
        C. Schwanitz \inst{\ref{i:Sw}}, 
        L. Teriaca \inst{\ref{i:Ge}},
        M. J. West \inst{\ref{i:US}}, 
           \and
        A.~N.~Zhukov \inst{\ref{i:Be}, \ref{i:Ru2}}
          }
          
   \institute{Department of Physics, Faculty of Science, University of Zanjan, : University Blvd., Zanjan,  45371-38791, Zanjan, Iran \label{i:ZN}
   \and
   Observatory, Faculty of Science, University of Zanjan, : University Blvd., Zanjan,  45371-38791, Zanjan, Iran \label{i:OZN}
   \and
   Solar-Terrestrial Centre of Excellence – SIDC, Royal Observatory of Belgium, Ringlaan -3- Av. Circulaire, 1180 Brussels, Belgium \label{i:Be}
   \and
   Université Paris-Saclay, CNRS, Institut d’Astrophysique Spatiale, 91405 Orsay, France \label{i:Fr}
   \and
   Max Planck Institute for Solar System Research, Justus-von-Liebig-Weg 3, 37077 Göttingen, Germany \label{i:Ge}
   \and
   Department of Mathematics, Physics and Electrical Engineering, Northumbria University, Newcastle Upon Tyne, NE1 8ST, United Kingdom \label{i:UK}
   \and
     ETH-Z\"urich, H\"onggerberg campus, HIT building, Z\"urich, Switzerland \label{i:Sw1}
   \and
 Physikalisch-Meteorologisches Observatorium Davos,World Radiation Center, 7260, Davos Dorf, Switzerland \label{i:Sw}
   \and
   RCAAM of the Academy of Athens, 4 Soranou Efesiou Street, 11527 Athens, Greece \label{i:acad_at}
   \and
   UCL-Mullard Space Science Laboratory, Holmbury St. Mary, Dorking, Surrey, RH5 6NT, UK \label{i:UK2}
   \and
   Leibniz Institute for Astrophysics in Potsdam (AIP), 
   An der Sternwarte 16, 14482 Potsdam, Germany\label{i:Ge2}
   \and
   Centre for mathematical Plasma Astrophysics, Mathematics Department, KU Leuven,
   Celestijnenlaan 200B bus 2400, B-3001 Leuven, Belgium\label{i:Ru}
   \and
   Southwest Research Institute, 1050 Walnut Street, Suite 300, Boulder, CO 80302, USA\label{i:US}
   \and
   Skobeltsyn Institute of Nuclear Physics, Moscow State University, 119992 Moscow, Russia\label{i:Ru2}
   }

\date{Received; accepted}
 
  \abstract
   {Accurate detections of frequent small-scale extreme ultraviolet (EUV) brightenings are essential to the investigation of the physical processes heating the corona.  }
   {We detected small-scale brightenings, termed campfires, using their morphological and intensity structures as observed in coronal EUV imaging observations for statistical analysis.}
   {We applied a method based on Zernike moments and a support vector machine (SVM) classifier to automatically identify and track campfires observed by Solar Orbiter/Extreme Ultraviolet Imager (EUI) and Solar Dynamics Observatory (SDO)/Atmospheric Imaging Assembly (AIA).}
   {This method detected 8678 campfires (with length scales between 400 km and 4000 km) from a sequence of 50  High Resolution EUV telescope  (\HRI)  174 \AA~images. From 21 near co-temporal AIA images covering the same field of view as EUI, we found 1131 campfires, 58\% of which were also detected  in \HRI\ images. In contrast, about 16\% of campfires recognized in \HRI\ were detected by AIA. We obtain a campfire birthrate of 2 $\times$ $10^{-16}{\rm m}^{-2}{\rm s}^{-1}$. About 40\% of campfires show a duration longer than 5\,s, having been observed in at least two \HRI\ images.  We find that 27\% of campfires were found in coronal bright points and the remaining 73\% have occurred out of coronal bright points.  We detected 23 EUI campfires with a duration greater than  245\,s.  We found that about 80\% of campfires are formed at supergranular boundaries, and the features with the highest total intensities are generated at network junctions and intense H~I Lyman-$\alpha$ emission regions observed by EUI/\HRIL. The probability distribution functions for the total intensity, peak intensity, and projected area of campfires follow a power law behavior with absolute indices between 2 and 3. This self-similar behavior is a possible signature of self-organization, or even self-organized criticality, in the campfire formation process.}
   {}
  \titlerunning{Automatic detection of campfires}
   \authorrunning{N. Alipour et al.}

   \maketitle
%

\section{Introduction}
Brightenings in the extreme ultraviolet (EUV) have been observed in the solar atmosphere for many years \citep{Vaiana1973ApJ,Vaiana1973SoPh, Golub1974ApJ,alipour2015, Madjarskacoronal2019,Shokri2021}. The smallest of these brightenings in the quiet Sun (QS) were measured by \cite {Berghmans2021} in 174 \AA~ images taken by the  Extreme Ultraviolet Imager \citep [EUI,][]{Rochus2020} on board the Solar Orbiter mission \citep{muller2020}. These tiny EUV brightenings, coined "campfires" in the quiet Sun that have been found to date are short-lived events in the low corona with structures that take the form of small-scale loops, elongated loops, loop apexes, and contact points between loops \citep{Berghmans2021,Mandal2021A&A}.  These campfires have length scales ranging from 400 to 4000 km with typical lifetimes of less than 200\,s. Campfires reach temperatures peaking at about 1 MK \citep{Berghmans2021} and occur at heights of 1 to 5 Mm above the Sun's visible surface \citep{Zhukov2021}.

Campfires may be rooted in low-lying magnetic structures, and magnetic reconnection might be their formation mechanism, however this has not yet been shown beyond reasonable doubt. Hence, one aspect to explore is to check for similarities to other flaring events, for instance, large flares, microflares, and nanoflares \citep{Parker1988ApJ}. \citet{Panesar2021ApJ} studied the magnetic properties of 52 campfires and found that most of them occur above magnetic polarity inversion (i.e., "neutral") lines with a  magnetic flux cancelation rate of $10^{18}$ Mx hr$^{-1}$. These authors concluded that magnetic flux cancelation could be the primary mechanism in the formation of campfires. Also, they estimated the magnetic energy for the system of campfires in the range $10^{26}$ to $10^{27}$ erg, which might be sufficient to locally heat the solar corona depending on their occurrence frequency. \cite{Chen2021} used a 3D magnetohydrodynamic simulation with MURaM code and came to the conclusion that magnetic reconnection may well be the primary driver of most campfires. They also proposed that campfires can significantly contribute to quiet-Sun coronal heating.  

\cite{Chitta2021} recently applied an automatic method based on different intensity thresholds to Atmospheric Imaging Assembly (AIA) images to identify such tiny flare-like phenomena. These authors estimated about 100 small events emerge per second on the entire Sun, which is insufficient to supply the required energy for quiet Sun coronal heating. Coronal bright points (CBPs) are larger brightenings that include small-scale loops in the low corona with X-ray or EUV emissions \citep{Madjarskacoronal2019}. \cite{alipour2015} studied the statistical properties of the CBPs in EUV emissions observed by AIA during 4.4 years. They calculated an average size of 130 Mm$^{2}$ and a lifetime ranging from a few minutes to several days. \cite{Hosseini2021} calculated the  energy loss of CBPs. They also showed that by extending the CBPs' energy to nanoflares ($> 10^{24}$ erg), the contribution of small-scale bursts might increase significantly, along with their importance for heating the corona.

Due to the variety in morphology, structure, and intensity of campfires, along with their large numbers, automatic detection methods need to be developed to identify, track, and determine the occurrence frequency of these events that may play a role in coronal heating. We applied an automatic detection algorithm for campfires by analyzing various properties. In particular, we used the Zernike moments \citep[ZMs;][]{khotanzad1990} and the support vector machine \citep[SVM;][]{hsu2011bayesian} to identify campfires from both \HRI\ at 174\,\AA~and AIA at 171\,\AA\  observations. The first few low-order ZMs describe the centroid of an image and its orientation; however,  higher order ZMs provide information about geometry and structures (lines, shapes, etc.) of the objects \cite{alipour2019}. Since ZMs define mapping an image to the orthogonal complete set of functions (Zernike polynomials), the correspondent moments are unique and independent features. Therefore, these particular properties for ZMs demonstrate that ZMs are suitable for training a classification machine.  A sub-image with various sizes may contain a campfire or not, thus, the SVM as a mainly double-class classifier is applied to recognize campfire and non-campfire sub-images.  

This work is organized as follows: Section \ref{data} gives an overview of data analysis. The automatic identification and tracking algorithms are explained in Section \ref{methods}. The statistical analysis and results are presented in Section \ref{res}. Finally, the work is summarized and concluded in Section \ref{conc}. 

\begin{figure*}[!htb]
\begin{center}
\hspace*{-2.cm}
\includegraphics[width=1.\textwidth]{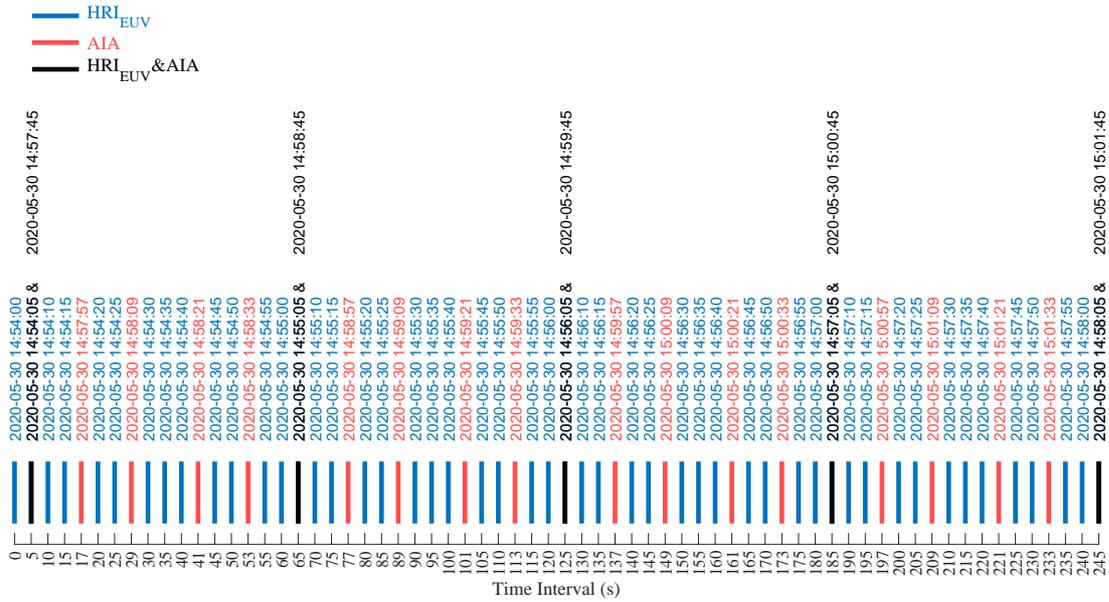}
\caption{Observation times  for \HRI\ (blue lines), AIA (red lines), and simultaneously EUI and AIA (black lines). The first simultaneous image was observed by AIA at 2020-05-30 14:57:45 UTC and \HRI\ at 2020-05-30 14:54:05 UTC.}
\label{fig1}
\end{center}
\end{figure*} 
\section{Data}\label{data}
The EUI is an instrument onboard the Solar Orbiter mission \citep{muller2020} with a Full-Sun Imager (FSI) and two High-Resolution Imager (HRI) telescopes \citep{Rochus2020}. The \HRI\ telescope samples at 174 \AA\ and is sensitive to temperatures of 1 MK. The \HRIL\ telescope observes images in the hydrogen Lyman-$\alpha$ passband centered at 1216 \AA. To study the statistical properties of campfires, we used the sequence of 50 \HRI\ and \HRIL\ images\footnote{EUI Data Release 2:         https://doi.org/10.24414/z2hf-b008}
with a cadence of 5\,s during a 245\,s time interval from 30 May 2020. On 30 May 2020, the \HRI\ FOV covered the SDO line-of-sight regions \cite[][Figure 1 therein]{Berghmans2021}. During these observations, Solar Orbiter was located at a distance of 0.556\,AU from the Sun, which resulted in a pixel size of 198\,km. To compare the AIA \citep{Lemen2012} and \HRI\ images, we correct for the difference in light travel time of 220\,s for the two spacecrafts.  We analyzed 21 AIA images at 171\,\AA\ during the same observation period. Figure~\ref{fig1} represents the observation time interval for \HRI\ (blue lines),  AIA (red lines), and simultaneously EUI and AIA (black lines) from the first simultaneous (EUI and AIA) image at 2020-05-30 14:54:05 UT. The last simultaneous (EUI and AIA) image was observed at 2020-05-30 14:58:05 UT. We analysed the AIA observations to compare the campfires detected in both instruments. To do this, we re-mapped all images to the Carrington coordinate system. Following \cite{Berghmans2021}, to preserve the \HRI\ resolution, we re-sampled the images on a 2400$\times$2400 grid with a 0.0163 heliographic degrees pitch. Also, we used a  radius of the Carrington projection sphere of 1.004$R_{\rm Sun}$ to minimize the average parallax over the FOV between the features observed by \HRI\ and AIA \citep{Zhukov2021}. Also, we used the SDO Helioseismic and Magnetic Imager (HMI) continuum images (14:35 UTC to 15:10 UTC) coaligned with AIA images to determine the supergranular cell boundaries.
\section{Methods}\label{methods}
Recently, \cite{alipour2012} and \cite{alipour2015}  investigated the feature detection methods based on machine learning and image processing for solar events (CBPs and mini-dimmings) observed by SDO/AIA images. ZMs provide unique and independent information for a class of features. In particular, ZMs for an event include abstractions of image information that reflect information on overall shape, geometrical, and morphological structure. Also, due to the Fourier term in Zernike polynomials, the absolute value of ZMs is invariant under rotation \citep{khotanzad1990}. Applying the image normalization, the ZMs will be translational and scaling invariant. This is done by transferring an image $I(x, y)$ into new image $I (x/a + x_{\rm cm}, y/a + y_{\rm cm})$ using the image centroid ($x_{\rm cm}$,$y_{\rm cm}$) and with a proper scale factor ($a$) \citep{raboo2017}. These bits of information about the content of a type of feature extracted by ZMs help us decide whether a solar sub-image exhibits an event or a non-event. 

To detect CBPs from \HRI\ at 174 \AA, we applied the previous CBPs automatic identification and tracking method \citep{alipour2015} with some minor updates on the training set. Since \HRI\ at 174 \AA\ and AIA at 171 \AA\ campfires are tiny events, we develop an automatic feature detection method for campfires by collecting enough information on both campfires and non-campfires as the training database to feed into the SVM classifier. The automatic algorithm for campfires consists of two main parts: a joint campfire classifier (JCC:Sec \ref{classifier}) and a campfire detection method (Sec \ref{identification}).   
 
\subsection{Joint campfire classifier}\label{classifier}
We collected (mainly by visual inspection)  sub-images (for 700 events) that show signatures of tiny EUV brightenings in EUI images. We probed the EUI sequence images for events containing small-scale loops, elongated loops, loop apexes, and contact points between loops, collectively termed campfires, with lifetimes in the range of 10 to 200\,s and with length scales ranging from 400 to 4000\,km. For later convenience, brightening features with sizes larger than 4000\,km are classified as CBPs. We took about 100 faint and small campfires, which were well identified by the wavelet automated detection method \citep{Berghmans2021}. Also, by visual inspection, we collected about 600 brightening features exhibiting the characteristics of these campfires. We also collected 700 sub-images for non-campfire features in the training set, including regions without campfires, some regions of large coronal loops, and so forth. In the next step, we computed the ZMs for campfires (positive class) and non-campfire (negative class) sub-images with five different sizes of $K_j\times K_j$ pixels ($j=1\cdots5, K_1=13,~K_2=17,~K_3=19,~K_4=23$, and $K_5=25$)  for the maximum order number $P_{max}=5$. For each of these five sub-images with different sizes, the ZMs with $P_{\rm max}=5$ has  21 data points placed in  data set 1. We also generated  ZMs with $P_{\rm max}=8$ (includes 45 data points) and five box sizes for both positive and negative classes, which are located in data set 2. The main reason for using two different ZMs data sets is to ensure that the classifier can identify campfires with different data points. Also, the variety of campfire sizes is the reason for the use of sub-images of various sizes. 
 
\begin{figure*}[!htb]
\begin{center}
     \begin{subfigure}
         \centering
         \includegraphics[width=12.cm,height=7.cm]{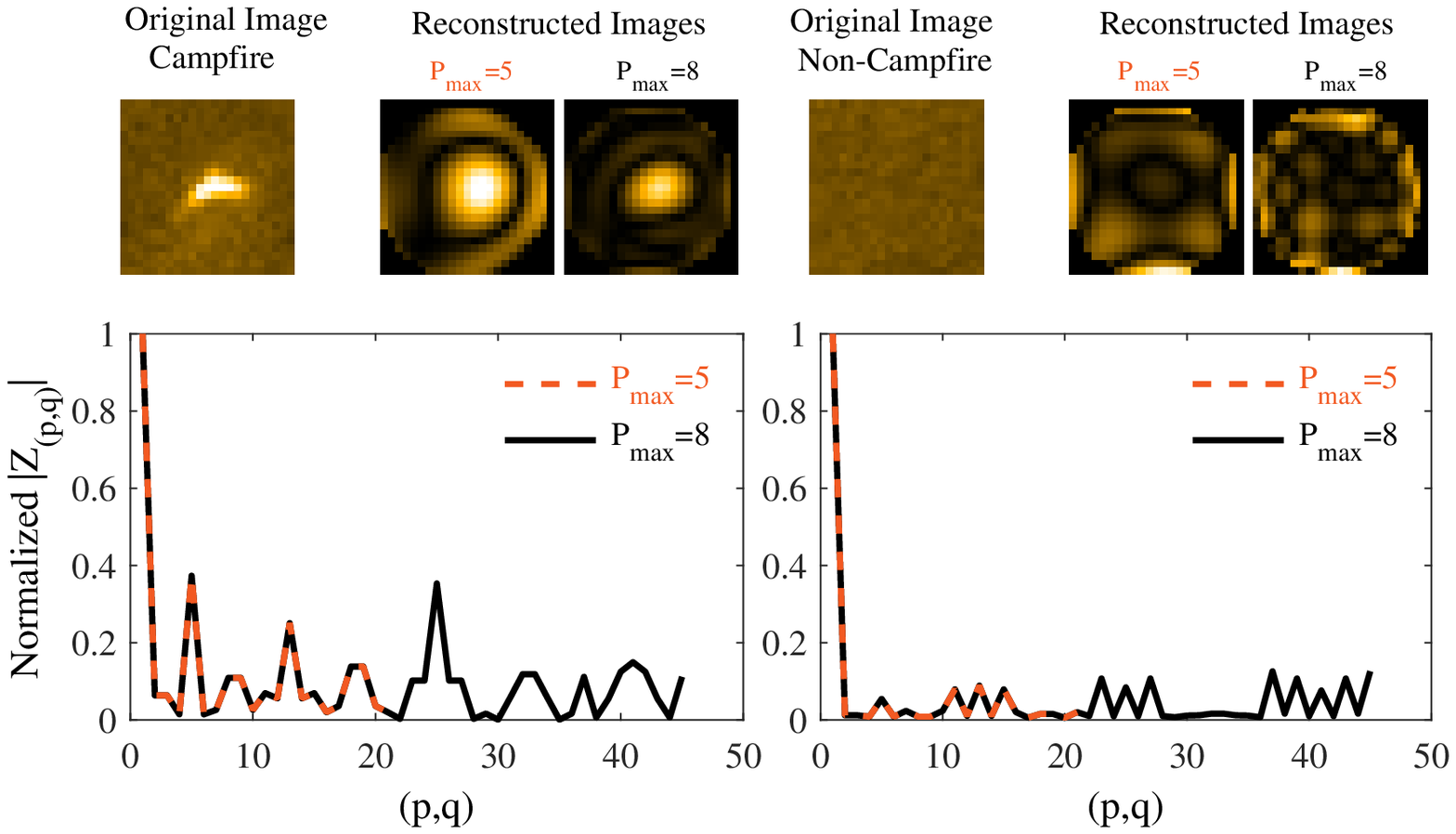}
     \end{subfigure}
     
     \begin{subfigure}
         \centering
         \includegraphics[width=12.cm,height=7.cm]{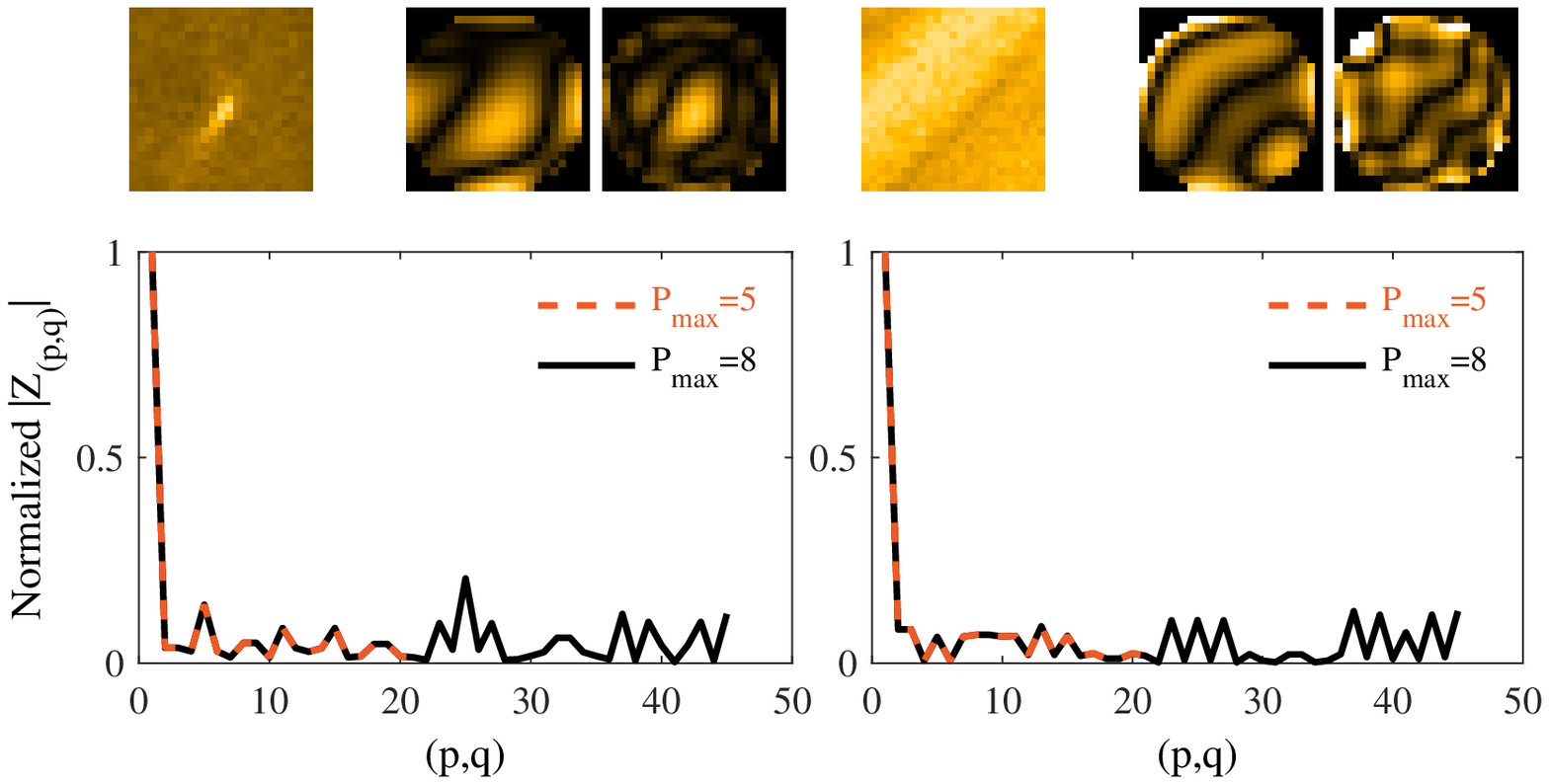}
     \end{subfigure}
\caption{Original \HRI\ images,  reconstructed images, and ZMs in relation to $P_{\rm max}=5$ (red dashed line) and 8 (black line) for two campfires (left panel) and two non-campfires (right panel). The order, $p,$ is a non-negative integer and repetition, $q,$ is an integer ($|q|-p\le0$ is an even number). For a given maximum order ($P_{\rm max}$), the  pair $(p,q)$ is labeled from 1 to $\sum_{p=0}^{P_{\rm max}} (p+1)$ \citep{alipour2019}. }
\label{fig2}
\end{center} 
\end{figure*}

Figure~\ref{fig2} demonstrates a campfire and a non-campfire sub-image (with a size of $23\times23$ pixels) in original \HRI\ and reconstructed images with ZMs concerning $P_{\rm max}=5$ (red dashed line) and 8 (black line). The figure shows that the ZMs for the campfire and non-campfire are distinguishable. For the ZMs of a given campfire, some structures (e.g., blocks) and variations look different from non-campfire ZMs. If we select another campfire sub-image with different sizes and then compute the ZMs, the block structure is slightly similar, with minor differences to the ZMs of other campfires because it is the structure and variations similarity that is registered in the moments. These differences between ZMs of campfires and non-campfires give us confidence in applying a machine classifier (SVM) to identify features from \HRI\ and AIA images. In other words, the ZMs of two classes of features (campfires and non-campfires) in multidimensional feature space (the dimension of ZMs) have enough distributable information for a  (statistically) learning machine to detect campfires.  

We set the SVM classifier with Gaussian kernel parameter (rbf\_sigma=15) in the training step to achieve a well-trained classifier. We randomly selected 70\% of both positive and negative classes (for each ZMs data set 1 and 2) to apply as the training set. The remaining 30\% of both classes was used as the test set. We used the same labels for both ZMs sets 1 and 2. In other words, we used two parallel classifiers for ZMs data set 1 (classifier 1) and ZMs data set 2 (classifier 2). We then compared the output of both classifiers. We consider a feature as a campfire if both classifiers have identified it in the positive class. Thus, we collected the respective labels for campfires and non-campfires in positive and negative classes as a joint campfire classifier (JCC) component. Analyzing the output of the JCC, we evaluated the performance of automatic identification for campfires and non-campfires.
To measure the performance of the campfire classification method, we applied various classification metrics. To do this, we use the elements of a confusion matrix (CM) in which the number of positive (P) and negative (N) features in the data set are terms of the CM \citep{powers2011}. The components of the CM are true positive (TP: campfires correctly recognized), false positive (FP: non-campfires incorrectly recognized), true negative (TN: non-campfires correctly recognized), and false negative (FN: campfires incorrectly recognized). The essential metrics are: precision, recall, f$_{1}$ score, accuracy, Gilbert Skill Score (GS), Heidke skill score (HSS$_{1}$, HSS$_{2}$), and the true skill statistic (TSS). The accuracy, precision, and f$_{1}$ are metrics for balanced class classifiers.  In the test set of balanced class classifiers, the number of features and non-features are approximately the same. The HSS and GS metrics are also defined for the balanced class classifiers.  However, the TSS is an essential metric to measure the performance of imbalanced class problems \citep{ Mason2010,Bloomfield2012,barnes2016, bobra2015,raboo2017, alipour2015}. Table \ref{tab1} summarizes the metrics' formulae  and their mean value for the joint classifier. The value of a given metric (e.g., TSS$>$0.8) shows that the classifier is well trained with acceptable performance to identify campfires.
\subsection{Identification and tracking of campfires}\label{identification}
\begin{figure*}[!htb]
\begin{center}
\hspace*{-3.cm}
\includegraphics[width=1.4\textwidth]{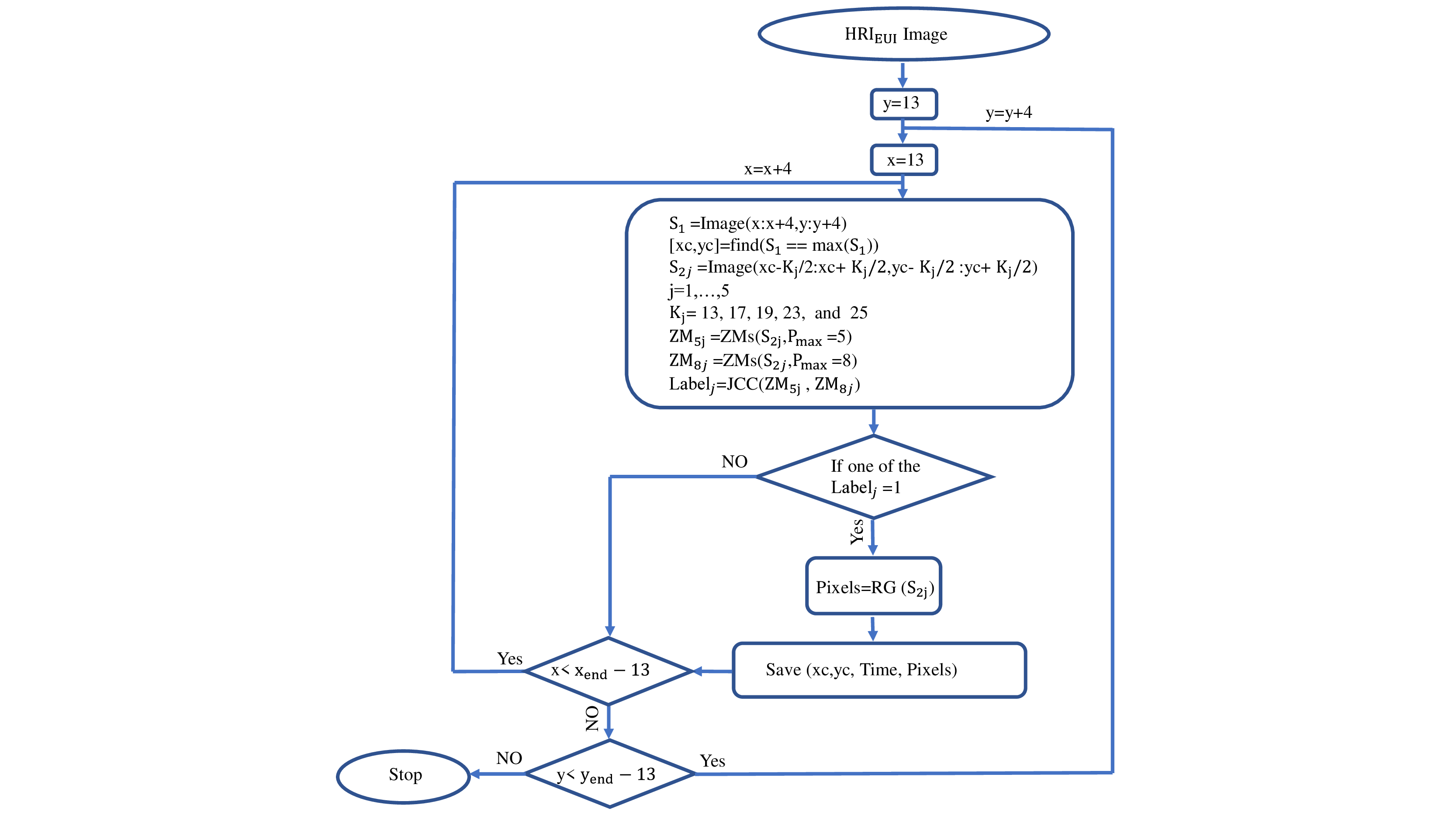}
\caption{Schematic flowchart for identification of \HRI\ and AIA  campfires . }
\label{fig3}
\end{center} 
\end{figure*}

We applied the JCC to identify campfires from EUI images.  
Figure~\ref{fig3} shows a schematic flowchart for the identification of campfires in each EUI image. The algorithm automatically scans a EUI image by a moving box. For each \HRI\ image, starting from $x = 13$ and $y = 13$, we extracted a small region with size $\Delta x=\Delta y=4$. We applied the joint campfire classifier to identify campfires from the EUI images. We determined the position, $x_c$, and $y_c$, of the maximum intensity inside the small region. This gives the location of the brightest pixel in the small region. In the next step, we selected larger region square boxes around the maximum intensity position with sizes $\Delta x\times\Delta y=K_j\times K_j~ (j=1\cdots5)$ pixels. The ZMs (for both $P_{\rm max}=5$ and 8) of these sub-images are computed. Then, the magnitudes of ZMs are fed to the JCC. The JCC picks up a label of 1 for a campfire candidate and 2 for a non-campfire sub-image. We applied a region growing (RG) function to the sub-image to extract the bright pixels. We consider the candidate a campfire if the linear length is greater than one pixel and smaller than 20 pixels. The image number (time) and output of the RG function for each campfire are saved. Then we move the small box first in $x$ direction up to the end of the grid and then in $y$ direction.

In the next step, we performed a tracking algorithm in the sequence of EUI images to obtain the duration of campfires. The algorithm tracks the identified campfires in the sequential \HRI\ images with an intersection of regions (joint pixels). To do this, we compared the locations (pixels) of the campfires at times $t_i$ ($i$ is the image number) and $t_{i+1}$ for all \HRI\ images. Next, we marked campfires with an intersection of regions (joint pixels in the sequence of images)  with the same labels. Then, using the labels and their related times, we computed the duration of campfires. 

\begin{figure}
\begin{center}
\includegraphics[width=0.5\textwidth]{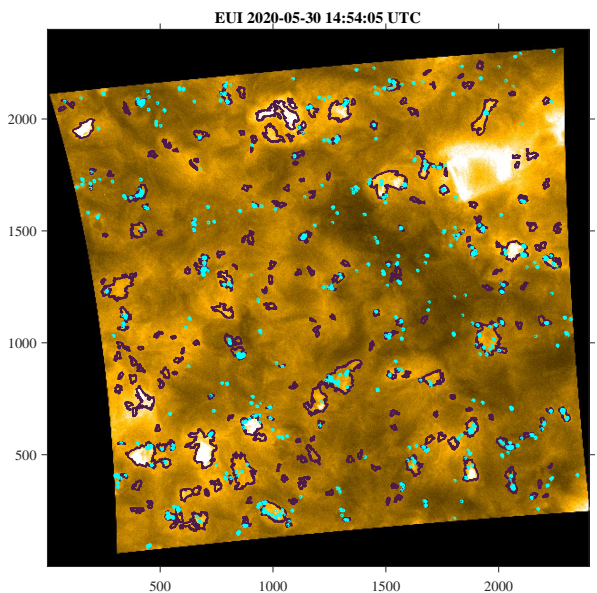}
\caption{ \HRI\ image at 174 \AA~ on 30 May 2020 14:54:05 UTC projected to Carrington coordinates. The code detected 240 coronal bright points (CBPs: purple contours) and 449 campfires (cyan contours). }
\label{fig4}
\end{center} 
\end{figure}

\begin{figure}
\begin{center}
\hspace{-2.cm}
\includegraphics[width=0.61\textwidth]{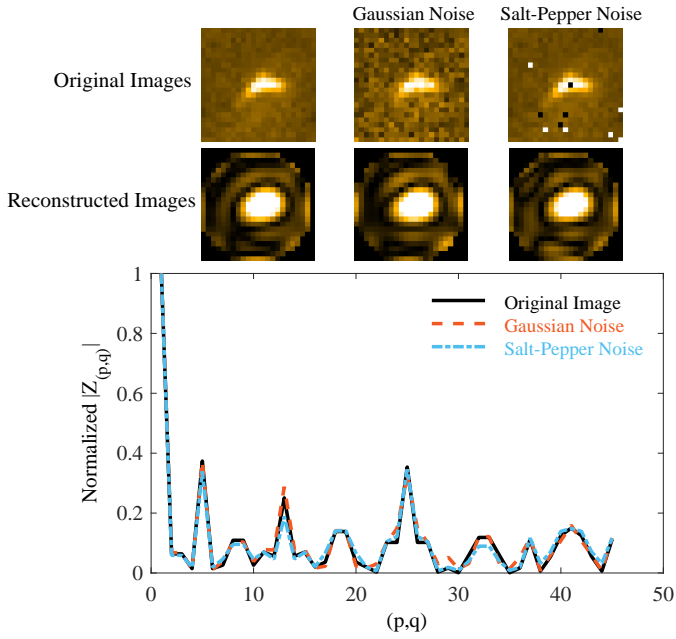}
\caption{Campfire at the original, artificial Gaussian, and salt-pepper noises (upper row) together with the reconstructed images (middle row) and their ZMs (bottom row). }
\label{fig5}
\end{center}
\end{figure}

\begin{figure}
\begin{center}
\includegraphics[width=0.5\textwidth]{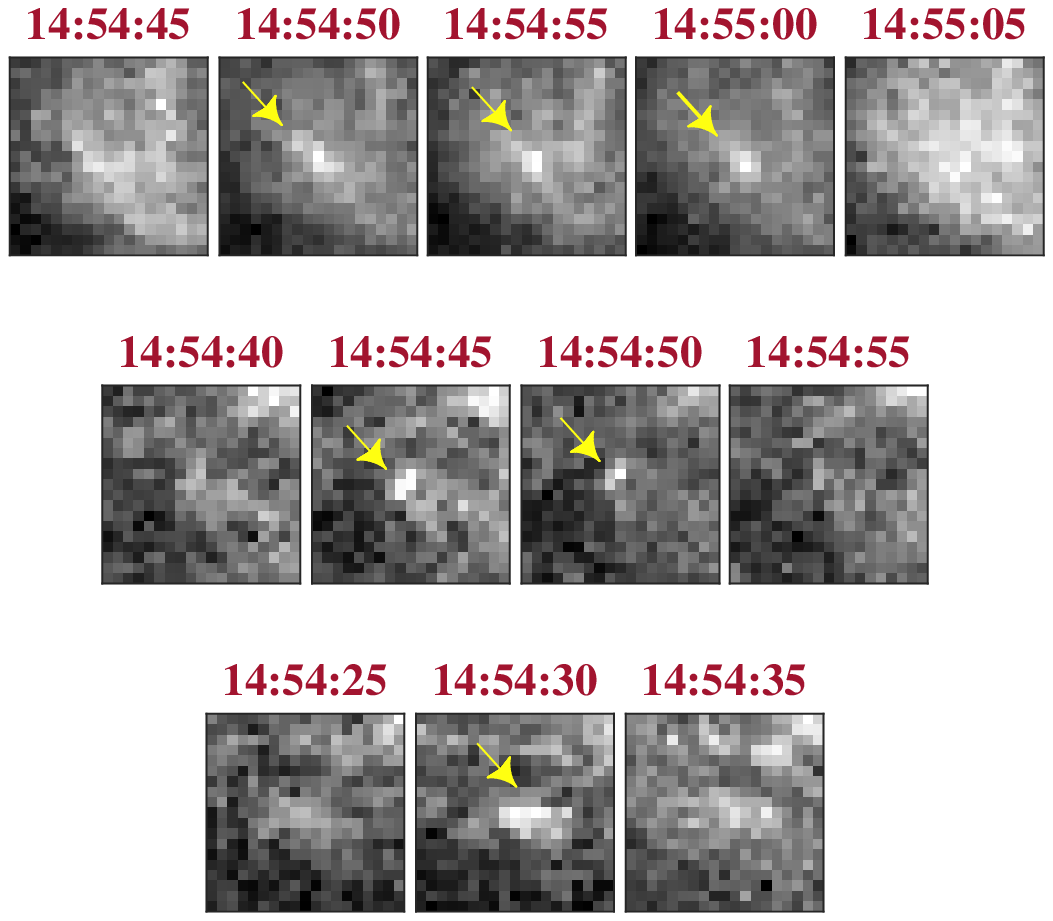}
\caption{ Three tracked samples with duration in the range of 15 s to 25 s (campfire 1, top row), 10 s to 20 s (campfire 2, middle row), and 5 s to 15 s (campfire 3, bottom row).}
\label{fig6}
\end{center}
\end{figure}
\section{Results}\label{res}
\subsection{\HRI\ brightenings}
We applied the automatic identification and tracking method for brightening features (CBPs and campfires) to the sequence of 50 Carrington projected \HRI\ images. Supplement S1 provides the sequence of 50 EUI at 174 \AA~ images  and the location of maximum intensity and time for the identified campfires as MATLAB structures. Supplement S2 gives movies for tracked campfires with the contour and label for each. 

Figure~\ref{fig4} represents an \HRI\ image with 449 detected campfires (cyan contours) and 240 CBPs (purple contours), within which 149 campfires were observed. We found that  27\% of campfires occurred inside CBPs. In other words, these campfires were hosted within CBPs and the remaining 73\% were observed out-of-CBPs. 

We detected 8678 campfires with length scales larger than or equal  ($\ge$) 400 km and smaller than 4000 km. This gives  a campfire birthrate of about 2 $\times$ $10^{-16}{\rm m}^{-2}{\rm s}^{-1}$. The number of events and their birthrate determined by the present method is approximately 5.4 times larger than those obtained using the wavelet detection scheme \citep{Berghmans2021}. Interestingly, the derived campfire birthrate is of the same order of magnitude as that found for explosive events \citep{Teriaca2004}, which required more investigation for the relationship between the two events. The wavelet detection scheme as pattern recognition is one dependent on the threshold used that detects a different number of events with changing the threshold value. After some minor modifications of the wavelet detection algorithm to work in space $(x, y)$ and time ($t$), the method was applicable for denoising images and identifying more campfires. However, the present campfires detection as a machine learning method identifies an event inside a sub-image with the specific characteristics in the feature space provided by ZMs. Figure \ref{fig5} shows a campfire at the original, artificial Gaussian, and salt-pepper noises together with ZMs and their reconstructed images. We find that ZMs are less sensitive to noise, as the reconstructed images are approximately the same for various types of noise. This is one of the reasons that we used ZMs instead of original images in feature space.   
We note that our training positive class contains the information of sub-images in which each sub-image has an event with a linear length greater than 400 km and smaller than 4000 km and lifetimes of more than 10\,s. This implies that we have chosen the suitable training features for the campfire class, excluding the noise features. However, the negative class may contain many non-campfire and noise features that do not have campfire characteristics.
Since the training feature space contains the morphology, structure, shape, and other information on the event, a detected single-frame event is considered a campfire if it consists of the bright pixel(s) and the neighboring pixels. In other words, for a sub-image that contains some individual or group of bright pixels without the characteristics of campfires in the feature space (ZMs), it will be classified as a non-campfire. 

\begin{table*}[bt]
 \renewcommand{\arraystretch}{1.6}
\vspace{0.5cm}
\caption{ Definitions of various skill scores as the measure for the performance of classifier and their mean values for campfires. } \label{tab1} \centering
 \begin{tabular}{p{3.6cm}@{\hspace{5mm}}cp{2.6cm}}
                \hline \hline
                Score & Formula & Campfires \\ \hline \hline
                Recall (positive)  & $\rm{recall}^{+} = \dfrac{\rm{TP}}{\rm{TP} + \rm{FN}}$ & $0.92\pm0.02$ \\  [2mm]\hline
                Recall (negative)  & ${\rm{recall}}^{-} = \dfrac{\rm{TN}}{\rm{TN} + \rm{FP}}$ & $0.93\pm0.02$ \\ [2mm]\hline
                
                Precision (positive) & ${\rm{precision}^{+}} = \dfrac{\rm{TP}}{\rm{TP} + \rm{FP}}$ & $0.93\pm0.02$ \\  [2mm]\hline
                Precision (negative) & ${\rm{precision}^{-}} = \dfrac{\rm{TN}}{\rm{TN} + \rm{FN}}$ & $0.92\pm0.02$ \\[2mm]\hline

                $f_{1}$ score (positive) & $f1^{+} = \dfrac{2 \times \rm{precision}^{+} \times \rm{recall}^{+} }{\rm{precision}^{+} + \rm{recall}^{+}}$ & $0.92\pm0.01$\\[2mm]\hline
                $f_{1}$ score (negative) & $f1^{-} = \dfrac{2 \times \rm{precision}^{-} \times \rm{recall}^{-} }{\rm{precision}^{-} + \rm{recall}^{-}}$ & $0.93\pm0.01$ \\[2mm]\hline

                Accuracy & $\rm{accuracy} = \dfrac{\rm{TP} + \rm{TN}}{\rm{TP} + \rm{FN} + \rm{TN} + \rm{FP}}$  & $0.93\pm0.01$ \\[2mm]\hline

                Heidke Skill Score (HSS1) & $\rm{HSS}_{1} = \dfrac{\rm{TP} - \rm{FP}}{\rm{TP} + \rm{FN}}$ & $0.85\pm0.02$ \\[2mm]\hline
                
                Heidke Skill Score (HSS2) & $\rm{HSS}_{2} = \dfrac{2 \times [(\rm{TP}  \times \rm{TN}) - (\rm{FN}  \times \rm{FP})]}{(\rm{TP} + \rm{FN}) \times (\rm{FN} + \rm{TN}) + (\rm{TN} + \rm{FP}) \times (\rm{TP} + \rm{FP})}$ & $0.85\pm0.02$ \\[2mm]\hline

                Gilbert Skill Score (GS)& $\rm{GS} = \dfrac{\rm{TP} - \rm{CH}}{\rm{TP} + \rm{FP} + \rm{FN} - \rm{CH}},$ \\[3mm] & $\rm{CH} = \dfrac{(\rm{TP} + \rm{FP}) \times (\rm{TP} + \rm{FN})}{\rm{TP} + \rm{FN} + \rm{TN} + \rm{FP}}$  & $0.74\pm0.04$ \\[2mm]\hline
                
                True Skill Statistic (TSS)&$\rm{TSS} = \dfrac{{\rm{TP}}}{\rm{TP} + \rm{FN}} - \dfrac{{\rm{FP}}}{\rm{FP} + \rm{TN}}$  & $0.85\pm0.02$ \\[2mm]\hline
        \end{tabular}
 \end{table*} 

Of the 8678 detected campfires, 3300 have lifetimes larger than 5 s and less than 245 s. We observed 23 campfires with a duration longer than 245 s. Figure~\ref{fig6} shows three tracked samples with lifetimes in the range of 15 s to 25 s (campfire 1, top row), 10 s to 20 s (campfire 2, middle row), and 5 s to 15 s (campfire 3, bottom row). Some factors such as identification errors (performance scores in Table \ref{tab1}), displacement of the bright pixels, and variation of the length scale for some campfires may affect the number of detected campfires. We found that the tracking algorithm suffers the false negative error in which the actual number of events is probably more than the detected campfires. Therefore, this affects the duration of detected events, and the actual lifetime for some campfires is more than the values mentioned above. So, the 3300 campfires with a duration of at least 5 s are the lower limit for this algorithm.   


\begin{figure}
     \centering
      \begin{subfigure}
         \centering
         \includegraphics[width=8.cm,height=6.cm]{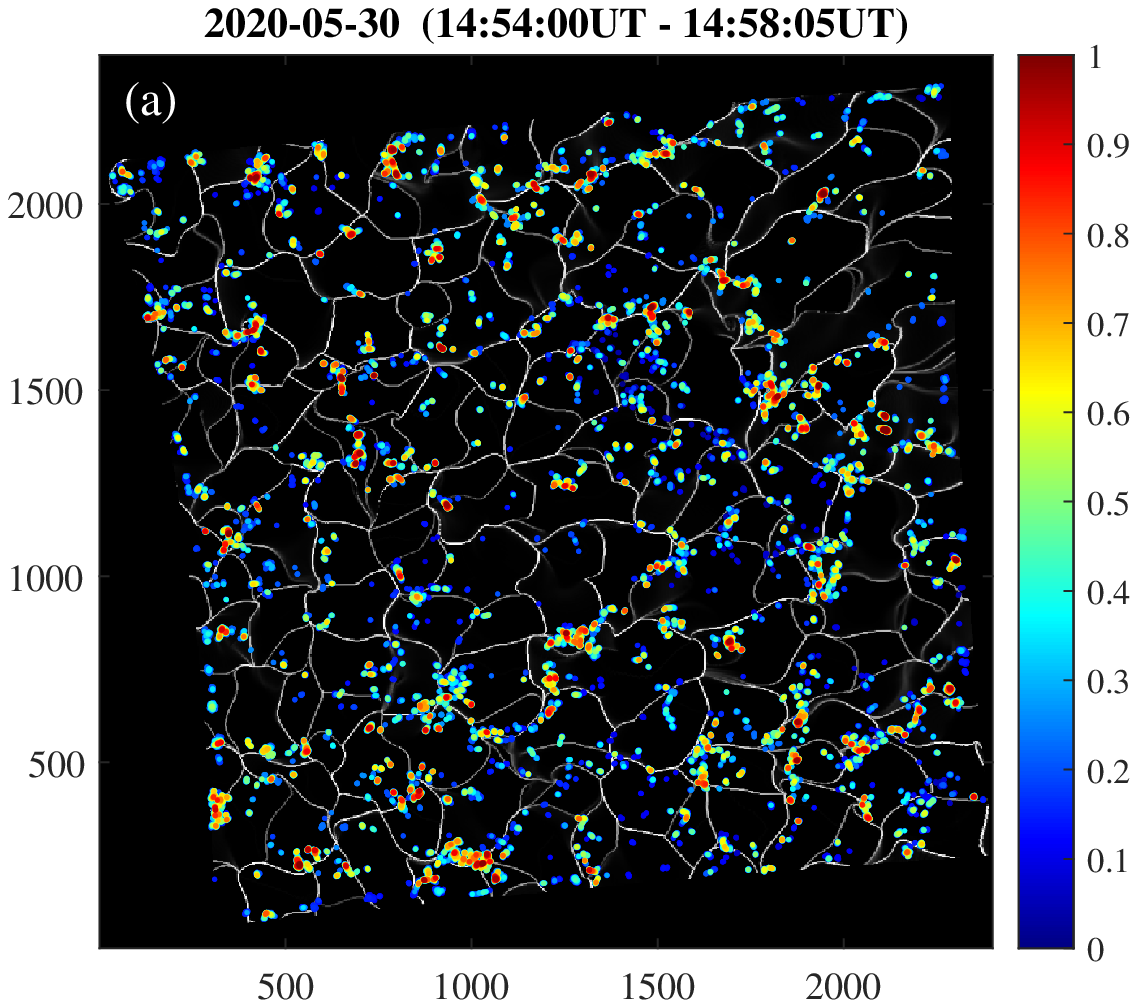}
     \end{subfigure}
     \begin{subfigure}
         \centering
         \includegraphics[width=8.cm,height=6.cm]{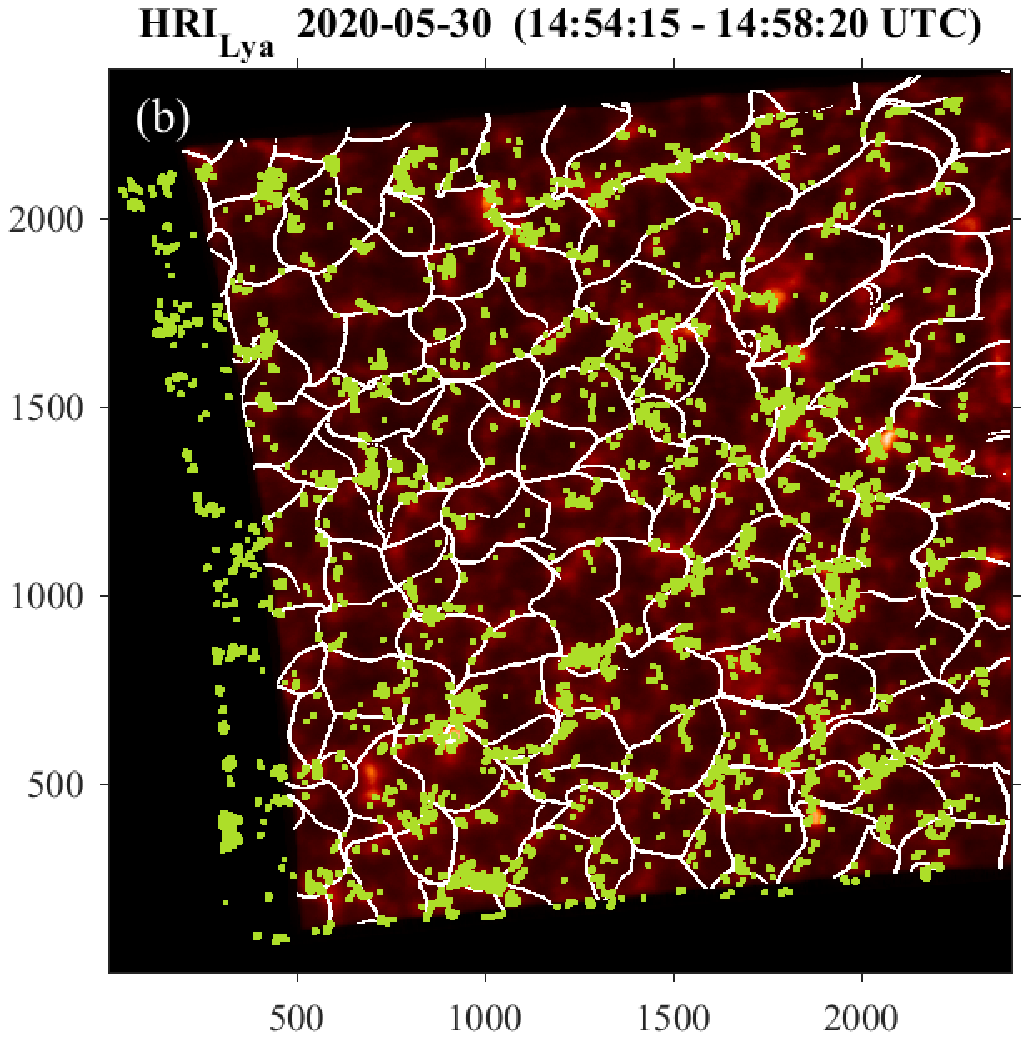}
      \end{subfigure}
     \begin{subfigure}
         \centering
         \includegraphics[width=8.cm,height=5.cm]{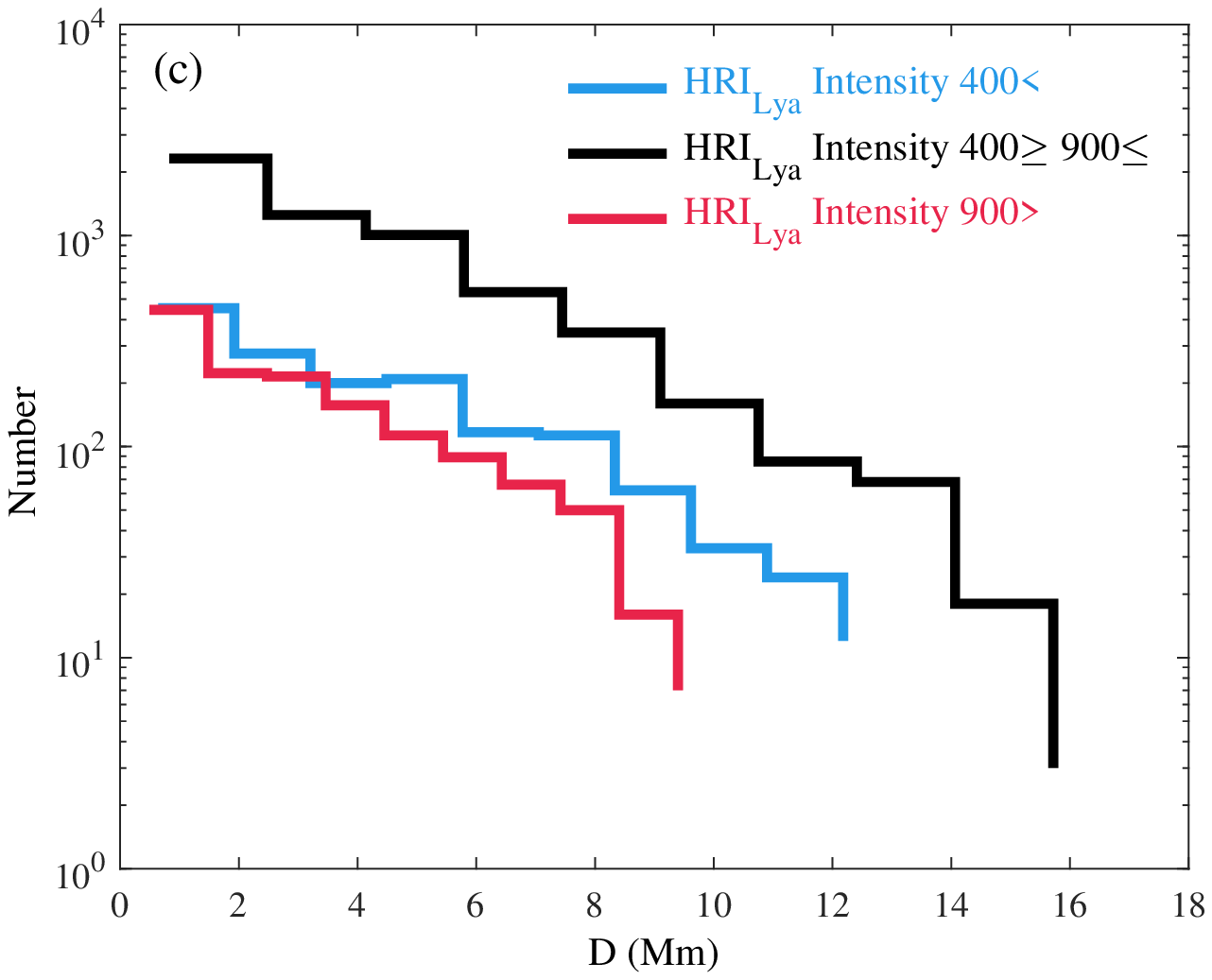}
     \end{subfigure}
     \caption{(a) Position of 8678 campfires (color dots: the total intensity of campfires) and supergranular boundaries (white lines).  The color map indicates the pixels' integrated intensity of campfires. (b) Average of 50 \HRIL images, supergranular boundaries (white lines), and position of campfires (green dots). (c) Frequency (number) of campfires' shortest distances (distance ($D$) of campfires centroid from the supergranular boundaries) for \HRIL intensity $<$400 DN/s (blue  line), 400 to 900 DN/s (black line), and $>$ 900 DN/s (red line). About 2/3 of campfires have the \HRIL  intensity of the centroid in the range of 400 to 900 DN/s.   }
\label{fig7}  
\end{figure}
 A campfire's intensity in \HRI\ is computed by integrating all corresponding pixel values at each image. The peak intensity is the maximum intensity during the observed campfire duration. The total intensity of an event obtains the summing of the intensity over the whole duration. 
Figure~\ref{fig7} shows the position of 8678 campfires, supergranular boundaries, an average of 50 \HRIL images, and the frequency of campfires' (centroid) shortest distances from the supergranular boundaries for \HRIL\ intensity. To determine the supergranular boundaries, we applied the ball tracking method on the respective SDO/HMI continuum images \citep{Potts2004,Attie2015,Attie2016} for tracking solar photospheric flows. The shortest distance of a campfire centroid (center of brightness in \HRI) from supergranular boundaries is computed.
We observed that campfires with a higher total intensity were mostly placed at supergranular junctions and high \HRIL\ emission regions. We found that most (80\%) of campfires are placed near supergranular boundaries (Fig.~\ref{fig7}a). In other words, about 80\% of the events have the shortest distances (minimum distance of a campfire centroid from the nearest supergranular boundaries) less than 5 Mm. The typical diameter of solar supergranules is significantly larger than 20 Mm \citep[e.g.,][]{DelMoro2004SoPh,Noori2019AdSpR}. About 2/3 of campfires have the \HRIL\  intensity of the centroid in the range of 400 to 900 DN/s (Fig.~\ref{fig7}b-c). We also found that the high \HRIL\ intensities were distributed around the supergranular boundaries (Fig.~\ref{fig7}b). The supergranular boundaries are the places for a high concentration of magnetic fields and converging photospheric flows \citep{eric}. As we observed, campfires are short-lived features similar to other types of small-scale transients: explosive events \citep[][]{Porter1991ApJ}, CBPs \citep[][]{yousefzadeh2016motion, Madjarskacoronal2019}, mini-coronal mass ejections \citep[][]{Honarbakhsh2016}, and blinkers \citep[][]{harrison1997euv,Shokri2021}. These are predominantly localized and analogous to those of flaring events (with a different amount of the involved energy). \citet{Panesar2021ApJ} showed that most of the 52 campfires in their investigation were above magnetic neutral lines with a  considerable rate of flux cancelation \citep{Ballegooijen1989ApJ}, which indicates that flux cancelation (i.e., an inherently resistive, magnetic reconnection mechanism) could be the primary mechanism in the formation of campfires. Therefore, it might be possible to hypothesize that campfires are flare-like events, but on a vastly smaller scale \citep[e.g.,][]{Giovanelli_1946, Dungey_1953}. This hypothesis needs to be investigated in further detail.

\begin{figure*}[!htb]
     \centering
      \begin{subfigure}
         \centering
         \includegraphics[width=13.cm,height=4.cm]{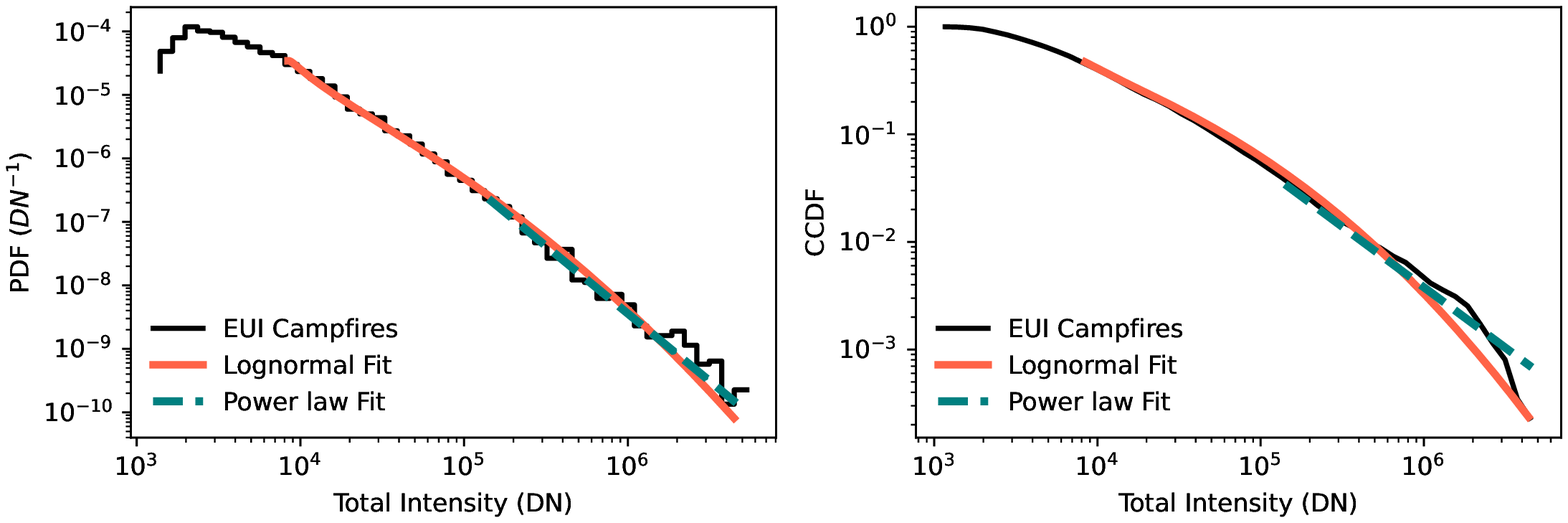}
     \end{subfigure}\\
     \begin{subfigure}
         \centering
         \includegraphics[width=13.cm,height=4.cm]{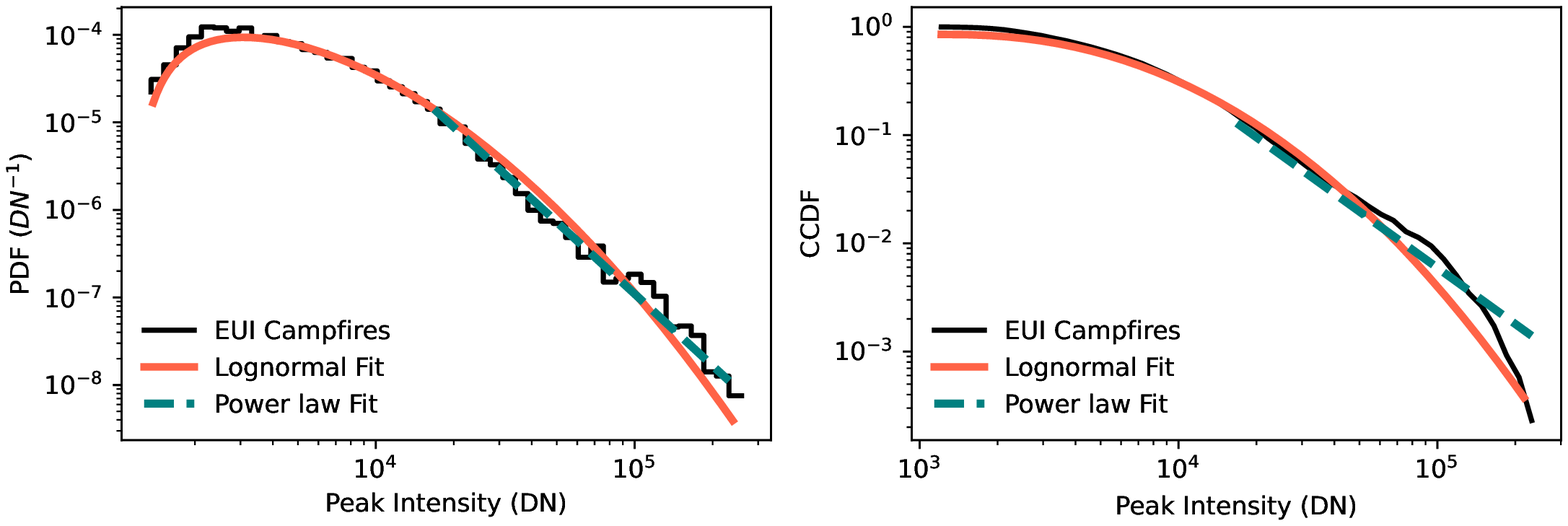}
      \end{subfigure}\\
     \begin{subfigure}
         \centering
         \includegraphics[width=13.cm,height=4.cm]{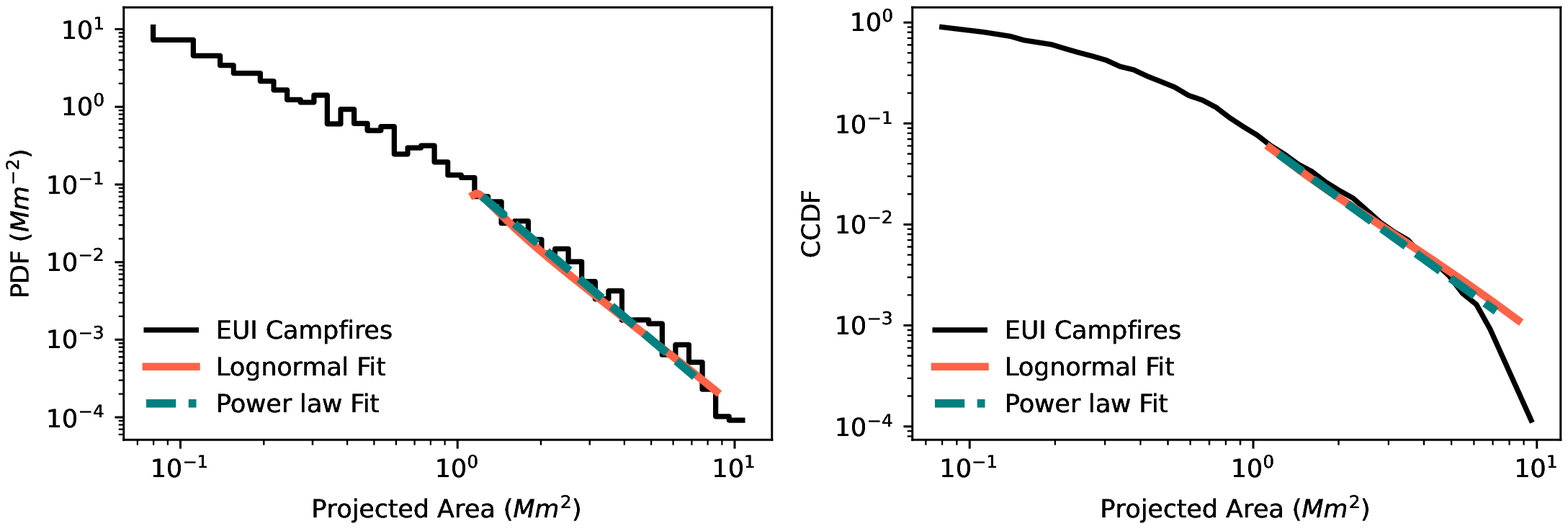}
     \end{subfigure}\\
     \begin{subfigure}
         \centering
         \includegraphics[width=13.cm,height=4.cm]{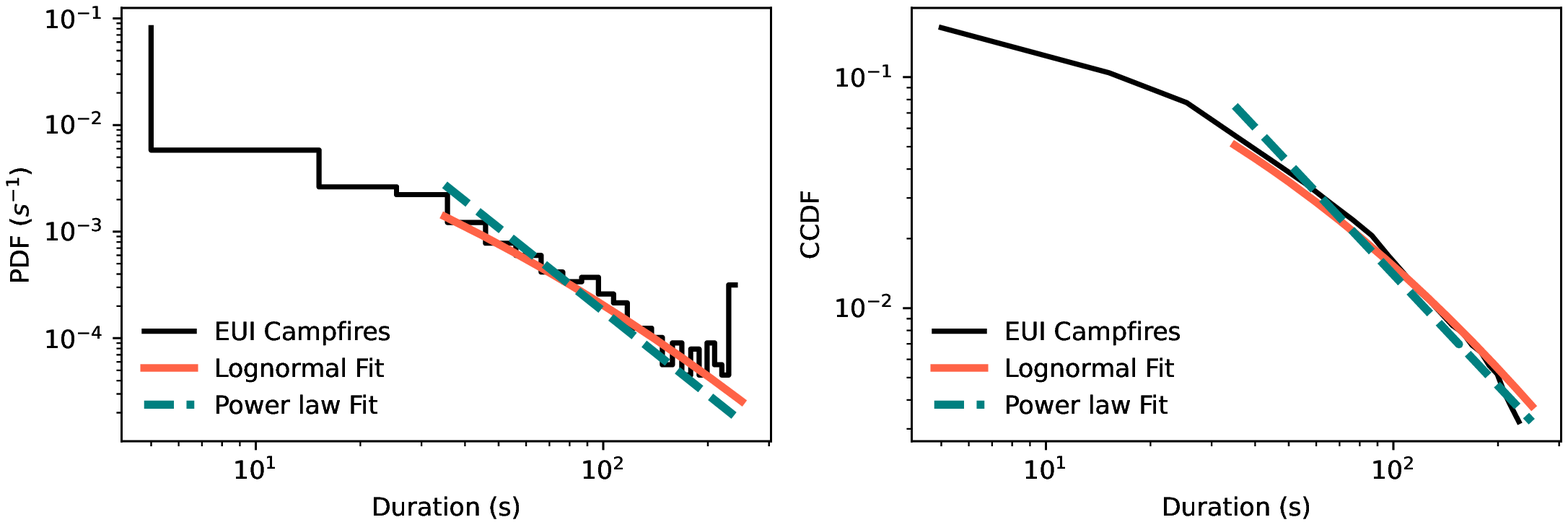}
     \end{subfigure}
      \caption{Probability distribution function (PDF; left column) and complementary cumulative distribution function (CCDF; right column) of the total intensity (first row), peak intensity (second row), projected area (third row), and duration (fourth row) of campfires observed by \HRI\ (black line). Power laws (power law fit: blue dashed lines) and lognormal functions  (lognormal fit: red line) are fitted to each distribution. }
\label{fig8}  
\end{figure*}

Figure~\ref{fig8} shows the probability distribution function (PDF) and complementary cumulative distribution function (CCDF) of the total intensity (first row), peak (maximum) intensity (second row), projected area (third row), and duration (fourth row) of campfires (black lines). To obtain the projected area for a campfire, we used every pixel area that a campfire occupies at some stage during its duration \citep{Berghmans2021}. As shown in the figure, the distributions show heavy-tailed behaviors. The power law and lognormal distributions are well-known distributions for modeling of such heavy-tailed flaring events \citep{Lin1984, crosby1993, krucker1998, parnell2000,Pauluhn2007A&A,Bazarghan2008A&A,klimchuk2009, Tajfirouze2012ApJ,farhang2019,Verbeeck2019ApJ}. The power law distribution function is
\begin{equation}
{\rm PDF}(x,x_{\rm min},\alpha)=\frac{\alpha-1}{x_{\rm min}}\left(\frac{x}{x_{\rm min}}\right)^{-\alpha},
\end{equation} 
where $x_{\rm min}$ is the cut-off and $\alpha$ is the power law index. To determine the minimum value for the cut-off ($x_{\rm min}$), we first create a power law fit that begins from a unique guess cut-off within the data. Then, the minimization of the distance between the data and power law fit guides to select the optimal value for $x_{\rm min}$ \citep{Clauset2009SIAMR}. The lognormal function is 
\begin{equation}
{\rm PDF}(x,\mu,\sigma)=\frac{1}{x\sigma\sqrt{2\pi}}\exp\left(-\frac{(\ln x-\mu)^2}{2\sigma^2}\right), 
\end{equation}
where $\mu$ is the mean and $\sigma$ is the shape parameter.

We applied a maximum-likelihood approach  \citep{Clauset2009} for fitting the power law and lognormal distributions to the total intensity, peak intensity, projected area, and duration for campfires. We applied a hypothesis test via Kolmogorov–Smirnov statistic that the null hypothesis supposes no significant difference between the observational distribution and the desired model. Instead, the alternative hypothesis supposes a significant difference between the observation and the model. Calculating a p-value, we can decide whether the model hypothesis is plausible for our data or not. A p-value smaller than  0.1  refutes the null hypothesis indicating that the specific model is ruled out \citep{mayo2006frequentist}. In contrast, we cannot refute the null for a p-value greater than  0.1. Using the bootstrapping approach, we determined the uncertainty and true value of the parameters for both the power law and lognormal distributions.  

\hspace{-0.5cm}
\begin{table*}[bt]
\caption{Parameters and p-values of fitted power law and lognormal distributions  for peak intensity, total intensity, projected area, and duration of campfires.  }
\centering 
\begin{tabular}{c   cc    ccc }
\hline \hline
\toprule
Dataset  & \multicolumn{2}{c}{Power law} &
\multicolumn{3}{c}{Lognormal} \\  \cline{1-6} 
& $\alpha$ & p-value & $\mu$ & $\sigma$ & p-value \\  
\cmidrule(l){2-3}       \cmidrule(l){4-6}
Total Intensity & $2.13\pm0.1$  &0.7 &  $9.40.\pm0.02$    & $1.78\pm0.01$  &0.2 \\
Peak Intensity  & $2.72\pm0.05$  &0.4 &  $8.72\pm0.01$     &$1.06\pm0.01$ &0.2  \\
Area Projected  &$3.02\pm0.12$ &0.2&  $-0.78\pm0.06$   &$1.34\pm0.04$  &0.2  \\
Duration        &$2.62\pm0.05$ &0.2&   $0.83\pm0.06$  &$1.53\pm0.07$  &0.8  \\
\hline
\bottomrule
\end{tabular}
\label{lab2}
\end{table*}

Table \ref{lab2} shows the p-values and fit parameters for power law and lognormal distributions (Fig.~\ref{fig8}) of the total intensity, peak intensity, area, and duration for the campfires. The p-values larger than 0.1 for both the power law and lognormal models mean we could not reject these models for the total intensity, peak intensity, projected area, and duration distributions. As expected, the lognormal distribution is a simpler explanation for most data because the power law distribution  only has a fit parameter ($\alpha$) and the lognormal model has two fit parameters ($\mu$, $\sigma$). However, in rare cases, a power law may be a better fit than a lognormal distribution if the data is inherently generated by mechanisms that obey a power law model. 
Multiplying random positive variables together is a simple mechanism for generating a lognormal distribution. However, the preferential attachment \citep{Gheibi2017ApJ} and  cellular automaton avalanche models \citep{Lu1993ApJ,farhang2019} are examples of generation mechanisms of a power law distribution. \cite{Pauluhn2007A&A} developed a model to simulate the nanoflare emissions observed by the Solar and Heliospheric Observatory (SOHO)/Solar Ultraviolet Measurements of Emitted Radiation (SUMER). In their model, an initial kick is generated by a power law distribution and evaluated by a multiplicative random process. The distribution of the resultant energies obeys a lognormal distribution, which saved the power law property in the memory of light curves. To determine the power law index for such light curves (e.g., small-scale emissions observed by SUMER and AIA), \cite{Bazarghan2008A&A, Tajfirouze2012ApJ, Upendran021ApJ}  developed methods based on neural networks.   

We obtained a power law index of about 2.1$\pm$0.1, 2.7$\pm$0.05,  3.0$\pm$0.05, and 2.6$\pm$0.05 for the total intensity, peak intensity,  projected area, and duration (Fig.~\ref{fig8}), respectively. The power law index (for peak intensity and total intensity of campfires as a class of small-scale events) greater than 2 is in agreement with the previous findings for energetic events \citep[e.g.,][]{krucker1998,Pauluhn2007,Bazarghan2008A&A, Tajfirouze2012ApJ,reale2014coronal}. Furthermore, the power law index of about 2.7 for the peak intensity distribution of campfires is in the range of indices obtained for the distribution of energy-loss flux of CBPs \citep{Hosseini2021}. Generally, however,  when deriving a power law index, there are error sources such as the instrument sensitivity, performance of a feature detection method (identification and tracking), estimation of modeling parameters, and the specific time of observations at a solar cycle \citep[e.g.,][]{parnell2000,parnell2002,Hosseini2021}. An accurate determination of the power law index for the energy distribution of flaring-like instabilities (from picoflares to large X-class flares) and their generating mechanisms is still an open problem \citep{Verbeeck2019ApJ}. Nevertheless, the power law behaviors may result from the self-similar features such as magnetic reconnection that generates campfires, similarly to magnetic field instabilities of self-organized criticality \citep{Lin1984, crosby1993,Aletti2000ApJ,Georgoulis2001A&A,Parenti2006ApJ,klimchuk2009, McAteer2016SSRv,Aschwanden2016SSRv, farhang2018}.

\subsection{Comparing AIA and \HRI campfires}

To compare the campfires observed in AIA 171 \AA~  images with the same FOV as \HRI\ 174 \AA~images, first we identified features in the original AIA images based on JCC. To do this, we used a moving box (Appendix \ref{classifier} and  \ref{identification}) with sizes slightly larger than half of the moving boxes for \HRI\ due to the different resolution of the two instruments. Then we remapped the images to Carrington coordinates.      
\begin{figure}
\begin{center}
\includegraphics[width=0.5\textwidth]{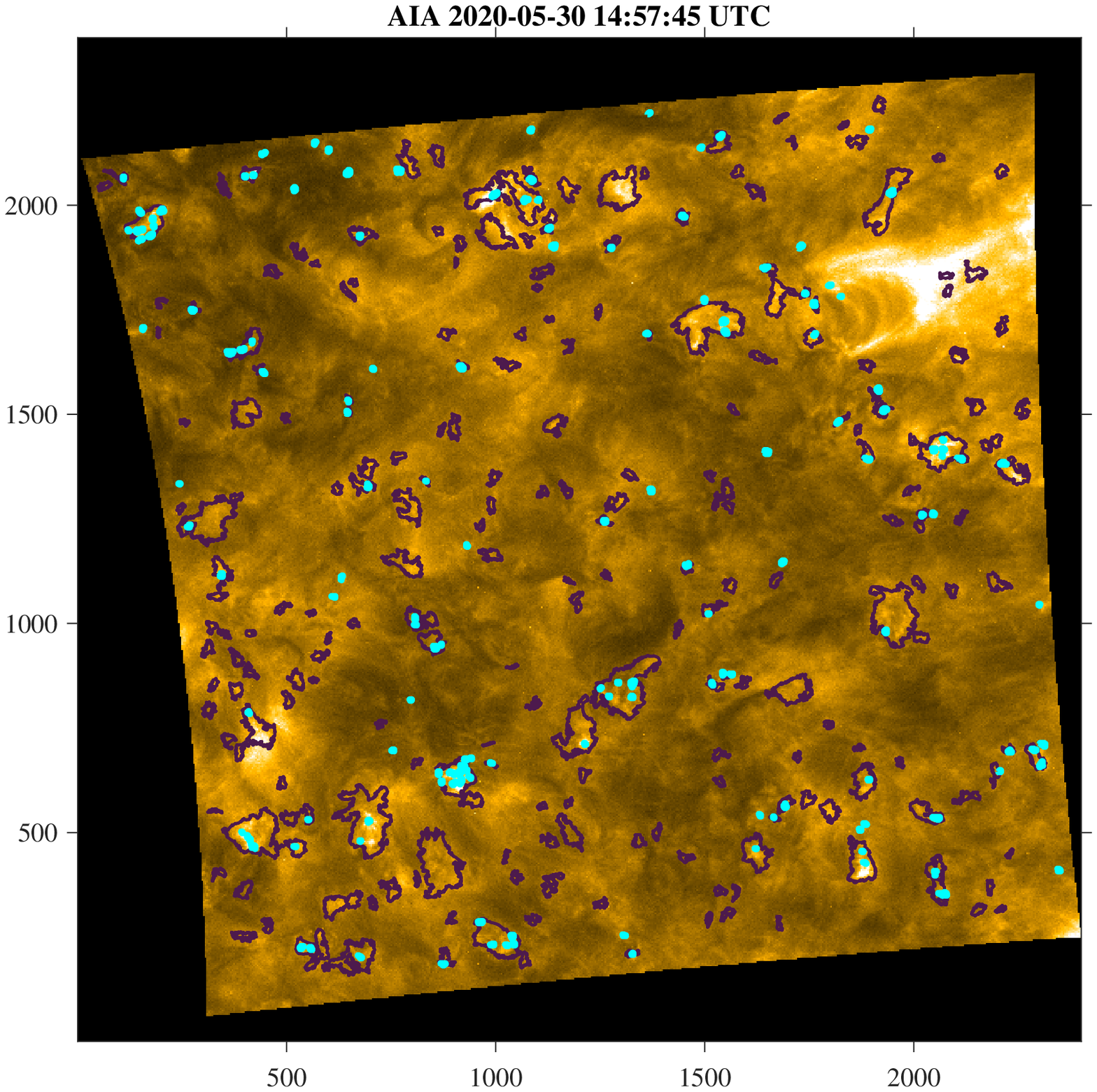}
\caption{ Positions of 147 campfires (cyan contours) in an AIA 171 \AA~image re-mapped to Carrington coordinates.  The positions of 240 CBPs (purple contours) are shown.}
\label{fig9}
\end{center}
\end{figure}
Fig.~\ref{fig9} shows an AIA 171 \AA~image (remapped to Carrington coordinates) with  147 campfires (cyan contours). At AIA's 12\,s cadence, 21 images are available during the \HRI\ image sequence (at the same FOV as \HRI). For AIA images, the method picked up 1131 campfires. As one output of the classifier, we observed that the campfire sub-images consist of at least two or more bright pixels (with linear length scale $\ge$880 km). We also observed that CBPs hosted about 48\% of the campfires. We observed about 500 campfires detected in at least a sequence of two AIA images.  Comparing the positions and times of features, we found that 50\% of the campfires observed in 21 AIA images were also detected in \HRI\ at 174 \AA~ images (e.g., Figs.~\ref{fig4} and \ref{fig9}). However, about 16\% of campfires detected in \HRI\ images were observed at AIA images. These differences in the number of campfires detected by AIA and \HRI\ may be related to the different spatial resolution due to the different heliocentric distances. At 0.556 AU from the Sun, EUI provides a spatial resolution of about 198 km, slightly less than half of AIA. These differences between the two analyses may be related to the slight differences in the AIA 171 \AA~ and \HRI\ 174 \AA~ bandpasses. A portion of these differences may relate to the different sizes of a feature in AIA and EUI, which are determined via a region-growing approach. In other words, some of the campfires detected in AIA with a length scale of about 4000 km have slightly larger length scales in EUI observations, so they were not considered as EUI campfires. Also, due to differences in the spatial resolution of the two instruments, we considered different lower limits on the length scale of campfires of about two pixels in both instruments (about 400 km at EUI and 880 km at AIA) to avoid the detection of the noise features.  
\begin{figure*}[!htb]
\begin{center}
\includegraphics[width=0.9\textwidth]{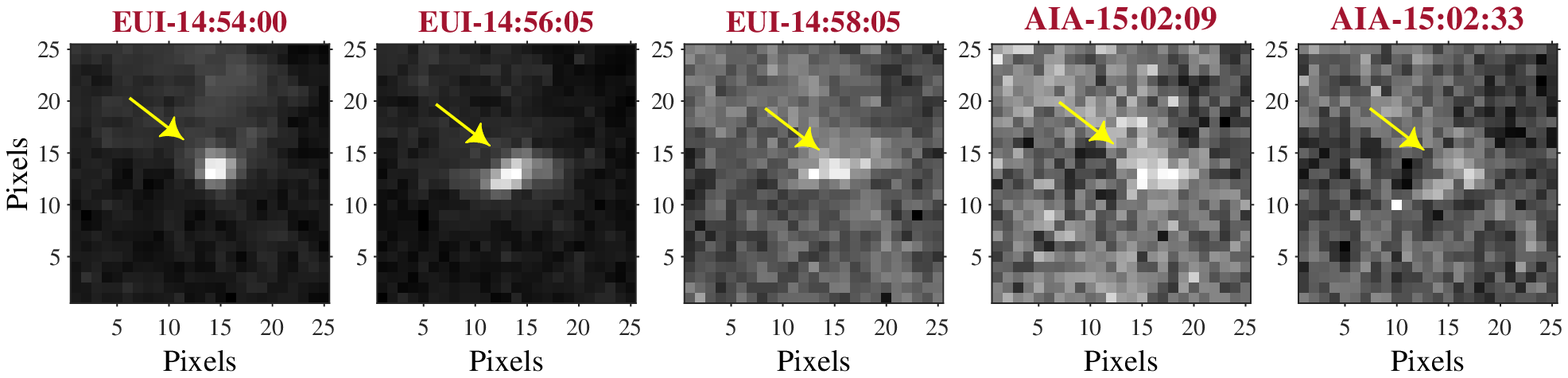}
\vspace{-9.01cm}
\caption{Campfire (supplement S3 movie, label 63)  was observed by both \HRI\  (position: x=737,y=1539 in Figure \ref{fig4}) and AIA (two latest sub-image at final stage). Its duration was more than 245 s in the EUI (14:54:00 to 14:58:05 UTC) observations. The actual duration in the range of 293 s to 312 s was observed in AIA images. The difference in light travel time of 220\,s needs to be considered between the two spacecraft.}
\label{fig10}
\end{center}
\end{figure*}

Figure~\ref{fig10} shows a campfire (supplement S3 movie) observed by \HRI\ (first three images) and AIA (two last images) from left to right. The duration of this campfire was greater than 245 s in the \HRI\ images re-mapped to Carrington coordinates with the same region as the AIA images. We detected this campfire in the AIA images with a duration greater than 293 s.   

\section{Conclusions} \label{conc}

In this work, we apply machine learning-based  automatic identification and tracking algorithms for CBPs and campfires to a sequence of 50 \HRI\ at 174 \AA~ and 21 AIA at 171 \AA~images recorded on 30 May 2020. The method uses information on features provided by ZMs that is fed into the SVM classifier to identify brightening features. The joint campfire classifier (JCC), as the combination of several SVM classifiers, uses a variety of information extracted with different moving box sizes and two maximum order numbers ($P_{\rm max}=5$ and 8) for ZMs to detect campfires from both  \HRI\ and AIA images. We measured the performance metric of the JCC machine (Table \ref{tab1}), which shows that the machine is well trained to identify the campfires. The tracking algorithm applied the RG function to extract bright pixels and followed events with the joint pixels through the sequence of images. 

We detected 8678 campfires with a birthrate of about 2 $\times$ $10^{-16}{\rm m}^{-2}{\rm s}^{-1}$ in \HRI\ images. About 3300 campfires have a duration longer than 10\,s, among which 23 campfires have a duration longer than 245 s (the last point at the tail of the duration PDF in Fig.~\ref{fig8}). Due to the false-negative error (i.e., erroneously identifying some of the campfire features as non-campfires) of the JCC machine for campfires, the 3300 events with a duration of more than 10\,s is a lower limit for the detection algorithm. It may well be possible that the actual number of campfires and their birthrate are higher than those identified by the present algorithm. In addition, CBPs hosted about 27\% of campfires and the rest occurred out of CBPs. By applying the JCC machine to 21 AIA images (the same FOV of \HRI\ images), we observed 1131 campfires with the linear length scale approximately in the range of 880 km to 4000 km. We observed that 500 campfires have a duration in the range of 12 s to 264 s. We compared the AIA and \HRI\ features (e.g., Figs.~\ref{fig4} and \ref{fig9}) that show about 50\% of the campfires observed in AIA were also identified in \HRI. In contrast, about 16\% of campfires recognized in \HRI\ were detected in AIA data.

 Due to the differences in the spatial  resolution of the two instruments, the tiny \HRI\ brightenings may appear weaker and fuzzier (i.e., very close to the background oscillations) in the AIA  observations; hence, we presume that they were not detected by this classifier machine. 
 This may imply that achieving observations of higher spatial resolution (i.e., closer Solar Orbiter perihelia) will give rise to (perhaps even significantly) more campfire detections. Another reason may be related to the differences of 174 \AA\ of \HRI\ and 171 \AA\ of AIA observations.
\citet{Tiwari2019} observed Hi-C brightening features for active regions that were bigger than \HRI\'s campfires. \citet{Joulin2016} and \citet{Chitta2021} studied the quiet-Sun small-scale brightening EUV events with SDO/AIA. The present method may provide  accurate detections of smaller brightenings for \HRI\ at heliocentric distances < 0.3 AU, so that the role of these events in the dynamics and, possibly, heating of the solar corona may be investigated in greater detail\citep{Berghmans2021}.

Using a maximum likelihood estimation, we determined a power law index of about 2.1$\pm$0.05, 2.7$\pm$0.04, 3.0$\pm$0.05, and 2.6$\pm$0.05, respectively, for the distribution of the total intensity, peak intensity, projected area, and duration. The power law behavior is one of the primary characteristics of self-organized criticality systems such as flaring events  \citep{Lu1993ApJ,Vlahos1995A&A,Strugarek2014SoPh,McAteer2016SSRv,Aschwanden2016SSRv,Aschwanden2018SSRv,farhang2019}. However, there are no unique reports for the power law index of flaring events and the variety of power law indexes ranging from 1.5 to 2.16 \citep[e.g.,][]{Hudson1991SoPh,Georgoulis1998A&A,Veronig2002A&A,Aschwanden2012} and 1.5 to 2.7 \citep[e.g.,][]{krucker1998,Berghmans1999SoPh,Hosseini2021} for large flares (M- and X- class) and small-scale flares (nanoflares, microflares, etc.), respectively, have been widely reported. This self-similar (or scale-free) property may shed light on the triggering process of campfires. In this context, recent studies of \cite{Panesar2021ApJ} and \cite{2022arXiv220213859K} show that a majority of campfires are associated with the cancelation of magnetic flux in the photosphere. We showed that the majority of \HRI\ campfires were placed at regions of  intense \HRIL emission and supergranular boundaries (lanes and junctions). We determined the supergranular boundaries by applying the ball-tracking method on the SDO/HMI images (Fig.~\ref{fig7}). About 2/3 of campfires have their \HRIL\  intensity (for the centroid) in the range of 400 to 900 DN/s. These pieces of evidence, along with the accumulation of campfires above network boundaries and neutral lines, suggest that magnetic reconnection may be considered an essential mechanism in the formation of campfires as dynamically evolving episodes of energy release.

\begin{acknowledgements}
Solar Orbiter is a space mission of international collaboration between ESA and NASA, operated by ESA. The EUI instrument was built by CSL, IAS, MPS, MSSL/UCL, PMOD/WRC, ROB, LCF/IO with funding from the Belgian Federal Science Policy Office (BELSPO/PRODEX PEA 4000112292); the Centre National d’Etudes Spatiales (CNES); the UK Space Agency (UKSA); the Bundesministerium für Wirtschaft und Energie (BMWi) through the Deutsches Zentrum für Luft- und Raumfahrt (DLR); and the Swiss Space Office (SSO). The ROB team thanks the Belgian Federal Science Policy Office (BELSPO) for the provision of financial support in the framework of the PRODEX Programme of the European Space Agency (ESA) under contract numbers  4000134474, 4000134088, and 4000136424. This work is supported by Swiss National Science Foundation - SNF. P.A. and D.M.L acknowledge STFC support from Ernest Rutherford Fellowship grant numbers ST/R004285/2 and ST/R003246/1 respectively. S.P. acknowledges the funding by CNES through the MEDOC data and operation center. The authors also gratefully thank the anonymous referee for very useful comments and suggestions that improved the manuscript. 
    
\end{acknowledgements}
\bibliographystyle{aa}
\bibliography{ref} 

\end{document}